\documentclass{jetpl}
\twocolumn
\usepackage{color}
\usepackage{graphicx}
\lat

\title{Surface acoustic wave controlled charge dynamics in a thin InGaAs quantum well}

\rtitle{Surface acoustic wave controlled charge dynamics in a thin InGaAs quantum well \ldots}

\sodtitle{Surface acoustic wave controlled charge dynamics in thin a InGaAs quantum well}

\author{Florian J. R. Sch\"ulein$^{+,*}$, Jens Pustiowski$^{+,*}$, Kai M\"uller$^{\dag}$, Max Bichler$^{\dag}$, Gregor Koblm\"uller$^{\dag}$,\\ Jonathan J. Finley$^{\dag}$, Achim Wixforth$^{+,*}$, Hubert J. Krenner$^{+,*}$\/\thanks{e-mail: hubert.krenner@physik.uni-augsburg.de}}

\rauthor{Florian J. R. Sch\"ulein et al.}

\sodauthor{Sch\"ulein, Pustiowski, M\"uller, Bichler, Koblm\"uller, Finley, Wixforth, Krenner}

\address{$^+$Lehrstuhl f\"ur Experimentalphysik 1 and Augsburg Centre for Innovative Technologies (ACIT), Universit\"at Augsburg, Universit\"atsstr. 1, 86159 Augsburg, Germany
\\~\\
$^*$Center for NanoScience {\it CeNS}, Geschwister-Scholl-Platz 1, 80539 M\"{u}nchen, Germany
\\~\\
$^{\dag}$Walter Schottky Institut and Physik Department, Technische Universit\"at M\"unchen, 85748 Garching, Germany}
\dates{xx. xxxxxx xxxx}{xx. xxxxxx xxxx}

\abstract{We experimentally study the optical emission of a thin quantum well and its dynamic modulation by a surface acoustic wave (SAW). We observe a characteristic transition of the modulation from one maximum to two maxima per SAW cycle as the acoustic power is increased which we find in good agreement with numerical calculations of the SAW controlled carrier dynamics. At low acoustic powers the carrier mobilites limit electron-hole pair dissociation, whereas at high power levels the induced electric fields give rise to efficient acousto-electric carrier transport. The direct comparison between the experimental data and the numerical simulations provide an absolute calibration of the local SAW phase.}


\begin{document}

\maketitle

Over the past 15 years surface acoustic waves (SAWs) have proven to be a powerful and versatile tool to control the electronic and optical properties of semiconductor nanostructures at radio frequencies up to several gigahertz \cite{Wixforth:1989,Rotter:1999,Kukushkin:2009,Metcalfe:2010,Fuhrmann:2011}. One prominent example is the dissociation of photogenerated excitons and the transport of the dissociated electrons ($e$'s) and holes ($h$'s) at the speed of sound within the plane of a quantum well (QW)\cite{Rocke:1997a,Wiele:1998}. This peculiar effect arises from the band edge modulation induced by the SAW generated electric fields in piezoelectric materials and can be observed also in one-dimensional systems such as embedded quantum wires \cite{Alsina:2002} or isolated nanowires \cite{Kinzel:2011}. Until now, the dynamics of this SAW driven process have been studied mainly in the limit of low SAW amplitudes \cite{Alsina:2001,Alsina:2003}. However, for high SAW amplitudes, i.e strong piezoelectric fields, a transition from a regime at which the low mobility of $h$'s limits the efficiency of this process to a fully field-driven regime is expected \cite{Garcia:2004}. While for low SAW amplitudes optical emission is modulated at the fundamental SAW frequency $f_{\rm SAW}$, in the field-driven regime this periodicity is expected to double providing a characteristic fingerprint.\\
Here, we report on the direct spectroscopic study of SAW controlled emission modulation and its dynamics spanning the entire range of SAW amplitudes and resolving the transition from mobility limited (low amplitudes) to field-driven (high amplitudes) dissociation of photo-generated ($e$-$h$) pairs in a thin, disordered QW. We resolve a characteristic transition of the emission modulation and its dependence on the local SAW phase which we find in good agreement with numerical simulations of the underlying drift and diffusion current equations\cite{Garcia:2004}. We find that such complete SAW power and phase scans of the optical emission provide a direct \emph{calibration of the absolute local SAW phase}.

\begin{figure}[htbp]
	\begin{center}
		\includegraphics[width=0.95\columnwidth]{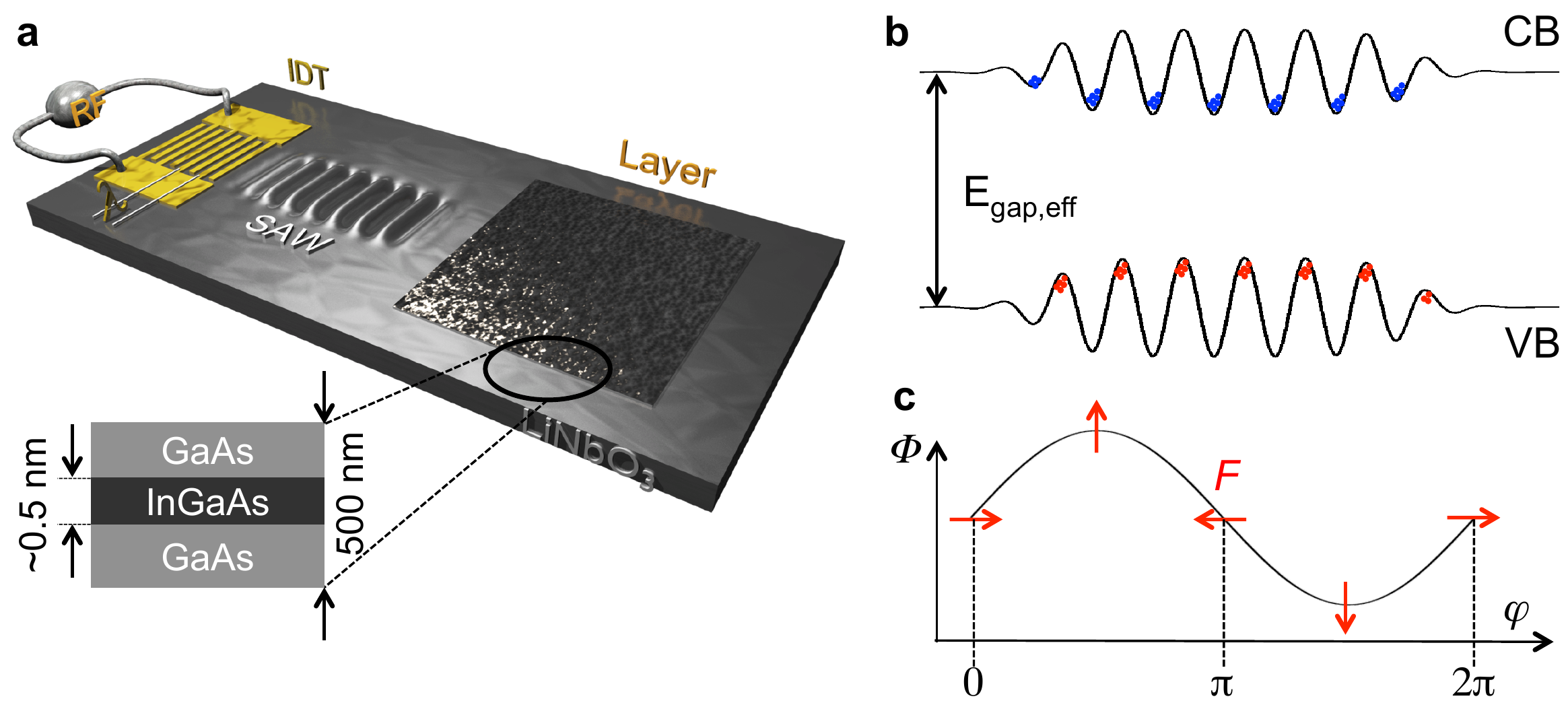}
		\caption{{\bf Figure 1} -- (a) Schematic of hybrid SAW device consisting of a ${\rm LiNbO_3}$ chip with an IDT to generate a SAW and an epitaxially transferred GaAs-based heterostructure. (b) Type-II band edge modulation induced by a SAW leading to a spatial separation of $e$'s and $h$'s. (c) Electric potential ${\it \Phi}$ (line) and electric field orientation $(F)$ at $(x=0)$ (arrows) as a function of the local SAW phase $(\varphi)$.}
		\label{fig:sample}
	\end{center}
\end{figure}
\begin{figure*}[htbp]
	\begin{center}
		\includegraphics[width=0.9\textwidth]{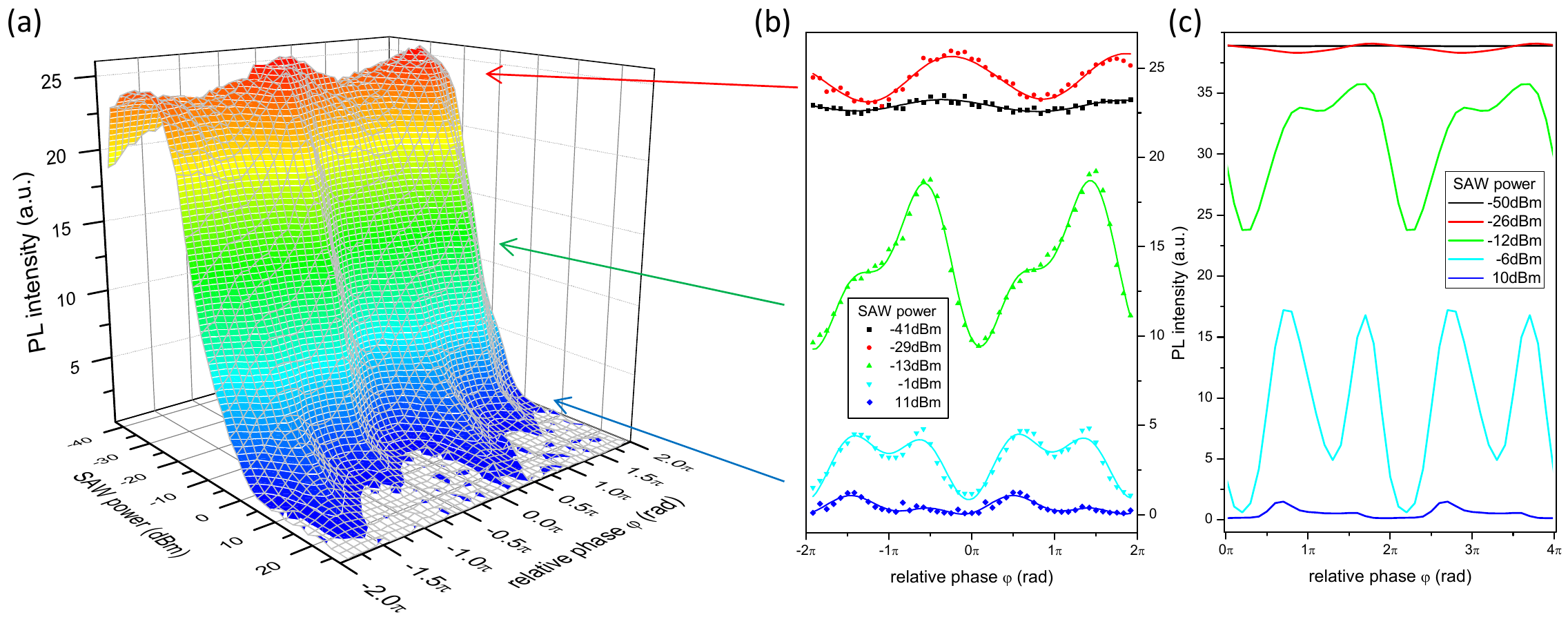}
		\caption{{\bf Figure 2} -- (a) Intensity map of the QW PL as a function of the acoustic power and relative phase. (b) Five representative phase scans over two SAW cycles $(-2\pi\leq\varphi\leq2\pi)$ at constant SAW powers (symbols) and fits of eq.\eqref{eq:intensity} (lines) demonstrating the transition from one maximum to two maxima per SAW cycle. (c) Simulated SAW intensity modulation (lines) corresponding to the experimental data.}
		\label{fig:results_3D}
	\end{center}
\end{figure*}
Our sample was grown by molecular beam epitaxy (MBE) on a semi-insulating GaAs (100) substrate. After growth of a GaAs buffer layer, we deposit a $100\:\mathrm{nm}$ $\mathrm{Al_{0.7}Ga_{0.3}As}$ sacrificial layer of for subsequent wet chemical etching. On top of this sacrificial layer, the optically active layer is grown which consists of a $220\:\mathrm{nm}$ GaAs buffer, $4\:\mathrm{ML}$ ${\rm In_{0.5}Ga_{0.5}As}$ and a $280\:\mathrm{nm}$ GaAs capping layer. During the growth of the the InGaAs layer, the substrate rotation was stopped to form a gradient of the In coverage which results in the formation of self-assembled quantum dots (QDs) in regions of high coverage and a thin wetting layer (WL) without QDs in the low coverage region\cite{Krenner:05a}. For the experiments presented here, we are exclusively focusing on material for which only a WL is present and no QDs are nucleating. Thus, we ensure that we resolve the SAW-dynamics of a thin and disordered QW formed by the InGaAs WL without carrier capture into fully confined QD levels superimposed. These layers are epitaxially lifted off the GaAs susbtrate by selective wet chemical etching using hydrofluoric acid in an established process \cite{Rotter:1999,Rotter:1997,Fuhrmann:2010} and transferred on a $128^\circ$ rot YX $\mathrm{LiNbO_3}$ host substrate. On this highly piezoelectric crystal SAWs can be excited all-electrically by interdigital transducers (IDTs) with a more than one order of magnitude increased efficiency compared to GaAs. A schematic of the final device is shown in Fig. \ref{fig:sample} (a) which allows for excitation of Rayleigh-type SAWs with a of wavelength $\lambda_{\rm SAW}=40\:\mathrm{\mu m}$ corresponding to a of frequency $f_{\rm SAW}=98\:\mathrm{MHz}$ ($T_{SAW}=10.2$\,ns). The strain and electric fields accompanying the SAW in the $\mathrm{LiNbO_3}$ substrate extend to the optically active semiconductor layers as the total thickness of the semiconductor film of $500\:\mathrm{nm}$ is $\sim80$ times smaller than the acoustic wavelength.
All optical experiments are carried out at $T=10\:\mathrm{K}$ in a liquid Helium flow cryostat in a conventional $\mu$-photoluminescence (PL) setup. For the optical excitation, we use an externally triggered pulsed laser diode emitting $\sim 60~{\rm ps}$ long pulses at $661\:\mathrm{nm}$ with $500\:\mathrm{nW}$ time averaged power. The laser beam is focused onto the sample by a $50\times$ objective to a $\sim2\:\mu\mathrm{m}$ diameter spot and the emitted PL is collected by the same objective. The PL signal dispersed by a $0.5\:\mathrm{m}$ grating monochromator is analyzed using a Si single photon avalanche diode (SPAD). In order to resolve the full SAW-driven dynamics of the PL even using time-integrated detection, we employ a phase-locked, stroboscopic excitation scheme \cite{Voelk:2011} for which we actively phase-lock an rf signal generator to excite the SAW and a pulse generator to trigger the laser diode by setting $n\times f_{{\rm laser}}=f_{{\rm SAW}}$ with $n$ integer. With this technique, we are able to tune the laser excitation over two full SAW cycles by tuning the \emph{relative} phase $-2\pi\leq\varphi\leq+2\pi$ of the rf signal which generates the SAW.\\

The SAW induced piezoelectric fields lead to a type-II band edge modulation in the QW which leads to a dissociation of photo generated $e$-hole pairs. At sufficiently high acoustic powers the two carrier species can be transferred to their respective stable points in the conduction and valence band which are separated by half the acoustic wavelength, thus suppressing their radiative recombination \cite{Rocke:1997a}. This is shown schematically in Fig. \ref{fig:sample} (b). At a fixed point in space, the electric potential ${\it \Phi}$ oscillates as the local SAW phase $\varphi$ tunes over $2\pi$. As depicted in Fig. \ref{fig:sample} (c), the resulting electric field $F$ gyrates with two characteristic maxima of its lateral component at $\varphi = 0,2\pi$ and $\varphi=\pi$ where the in-plane component of $|\nabla {\it \Phi}|$ is maximum. At these characteristic values of $\varphi$ \emph{field-driven} processes are expected to dominate \cite{Garcia:2004} giving rise to two minima of the QW emission per SAW cycle which become observable at high SAW amplitudes. 
\\

\begin{figure*}[htb]
	\begin{center}
		\includegraphics[width=0.9\textwidth]{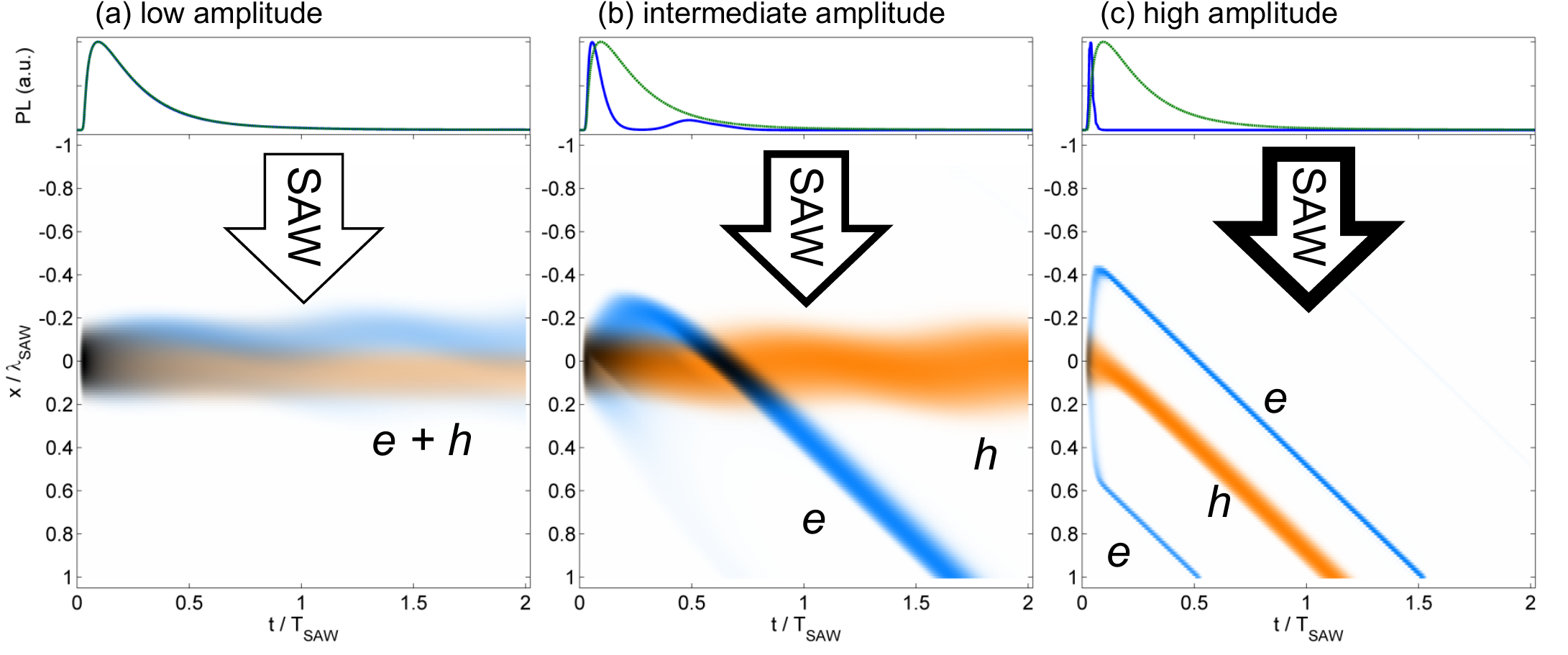}
		\caption{{\bf Figure 3} -- Simulated $e$ and $h$ spatial distributions as a function of time (lower panels) and the resulting temporal evolution of the QW emission (upper panels,blue solid lines) for low (a), intermediate (b) and high (c) SAW amplitudes. The unperturbed PL transient is plotted as a reference (dashed green lines). The optical excitation occurs at $\varphi=3\pi/2$ of the downward propagating SAW.} 
		\label{fig:simdensities}
	\end{center}
\end{figure*}
In Fig. \ref{fig:results_3D} (a) we present the measured PL emission intensity of the QW as a function of the RF power, -41\,dBm$ \leq P_{RF}\leq 27$\,dBm, applied to the IDT and the (local) SAW phase $\varphi$ tuned from $-2\pi$ to $+2\pi$. As the RF power is increased, the emission shows the characteristic quenching behavior due to the dissociation of $e$-$h$ pairs and the resulting inhibition of their radiative recombination. At fixed $P_{RF}$ the emission exhibits oscillatory behavior as $\varphi$ is tuned. At low $P_{RF}<-20$\,dBm, we observe a modulation with one maximum per SAW cycle (i.e. two maxima over the $4\pi$ range studied here). As $P_{RF}$ increases, a second oscillation with two maxima per cycle sets in and becomes more prominent at higher RF power levels. To further analyze this data we plot $\varphi$ scans at selected characteristic RF power levels in Fig. \ref{fig:results_3D} (b). In this data, we identify four regimes:\\ (i) For smallest acoustic powers below $-40\:\mathrm{dBm}$ (black line) no pronounced modulation is resolved.\\ (ii) At small SAW powers, $-40\:\mathrm{dBm}\leq P_{SAW}\leq-25\:\mathrm{dBm}$ (red line) we observe a weak $\lesssim10\%$ peak-to-peak modulation with the fundamental period of the SAW. Moreover, the overall intensity increases compared to no SAW excited. Here, the band edge modulation induced by the SAW is weak and, therefore, the finite mobilities of $e$'s and $h$'s are limiting the carrier transport. 
\\ (iii) In the intermediate power range (green and light blue lines), $-25\:\mathrm{dBm}\leq P_{SAW}\leq10\:\mathrm{dBm}$, two modulations of the emission are superimposed. In addition to the aforementioned modulation with \emph{one maximum}, a second oscillation with \emph{two maxima} per SAW cycle is clearly resolved. This additional modulation becomes more pronounced with increasing RF power pointing towards the predicted transition to fully field-driven carrier dissociation and transport. 
\\(iv) At the highest power levels accessible, $P_{SAW}\geq10\:\mathrm{dBm}$, field-driven processes dominate and the overall emission is almost completely quenched. Most notably, the two characteristic maxima per SAW cycle exhibit dissimilar amplitudes (dark blue).

In order to explain our experimental findings, we numerically solve the underlying drift and diffusion equations for $e$'s and $h$'s in the acoustically defined potential as a function of $\varphi$ and $P_{RF}$. For these simulations we assume an $e$ mobility of $(\mu_e=100\,{\rm cm^2/Vs})$, one order of magnitude lager than that of $h$'s $(\mu_h=10\,{\rm cm^2/Vs})$. In addition, we use the measured radiative lifetime of the QW excitons ($\tau = 0.6$\,{\rm ns}), the excitation laser spot diameter and its pulse duration as input parameters. The actual acoustic power is calculated from the applied RF power using the conversion efficiency of the IDT determined from its reflected RF power spectrum ($S_{11}$). In these simulations we first calculate the temporal and spatial evolution of carrier densities $e(x,t)$ and $h(x,t)$ and from the emission intensity given by $I(x,t)\propto e\cdot h$.
The such calculated integrated emission intensity is plotted as a function of $P_{RF}$ in Fig. \ref{fig:results_3D} (c) for five different acoustic powers corresponding to the approximate experimental power levels. Clearly, our simulations reproduce the two SAW-driven oscillations observed in the experimental data very well. This excellent agreement directly confirms that the observed spectroscopic signatures can be fully treated using a classical model of $e$'s and $h$'s in a SAW generated potential. To investigate the underlying processes in more detail, we plot the $e(x,t)$ (blue) and $h(x,t)$ (orange) in the main panels of Fig. \ref{fig:simdensities} and the resulting temporal evolution of the emission (upper panels) for three selected acoustic powers for an optical excitation at $\varphi=3\pi/2$. The two logarithmic color scales encoding $e$'s and $h$'s are added up. Thus dark gray corresponds to a large spatial and temporal overlap. (i+ii) For the case low acoustic powers [cf. Fig. \ref{fig:simdensities} (a)], corresponding to the red traces in Fig. \ref{fig:results_3D} (b) and (c), both carrier species are weakly affected in the externally induced potential. Due to their higher mobility $e$'s are "shaken" more effectively leading to a reduction of their temporal overlap with the nearly stationary $h$'s. Thus, the QW emission extends to longer times compared to the unperturbed decay without SAW (dashed line). This weak effect gives rise to a weak modulation of the overall intensity with one maximum per SAW cycle. (iii) For intermediate acoustic powers [cf. Fig. \ref{fig:simdensities} (b), green and light blue traces in Fig. \ref{fig:results_3D} (b) and (c)] $h$'s still remain stationary. However, $e$'s are efficiently and rapidly transferred to their stable points in the conduction band at which they are conveyed at the speed of sound by the SAW. This field-driven dissociation gives rise to a fast quenching of the emission (upper panel) which leads to a intensity modulation with two maxima per SAW cycle. Since $F\propto\sqrt{P_{RF}}$, this process and the resulting modulation become more pronounced for increasing $P_{RF}$. Moreover, the $e$'s initially transferred to the minimum located against the SAW propagation are conveyed across the stationary $h$'s exactly half of the acoustic period later. Thus, a time delayed emission maximum is clearly resolved in the calculated time transient in the upper panel of Fig. \ref{fig:simdensities} (b). (iv) For high acoustic powers [cf. Fig. \ref{fig:simdensities} (c), dark blue traces in Fig. \ref{fig:results_3D} (b) and (c)] both carrier species are efficiently conveyed by the SAW at their stable points. Due to the high electric fields the transfer to these points occurs quasi-instantaneously giving rise to a rapid temporal decay of the QW emission as seen in the upper panel of Fig. \ref{fig:simdensities} (c). Due to the rapid nature of the field-driven charge separation, emission is \emph{only} observed when either $e$'s and $h$'s are generated at their stable points. The different intensities of the two emission maxima  reflect the different mobilities of $e$'s and $h$'s. The higher maximum corresponds to $\varphi=\pi/2$ where $e$'s are excited at their stable point in the bandstructure and the charge separation is dictated by the lower hole mobility. In contrast, at $\varphi=3\pi/2$ where $h$'s are stationary and higher mobility $e$'s are transferred to their stable point which in turn gives rise to a faster charge separation and reduced emission intensity. Therefore, we are able to obtain an \emph{absolute} calibration of $\varphi$ from the intensity modulation measured at highest RF powers.\\ 

In order to analyze the evolution of the SAW modulation of the QW emission intensity we fit all experimental data sets using two oscillations given by
\begin{equation}
	I(\varphi) = I_0 + I_1\cdot \sin(\varphi-\gamma_1) + I_2\cdot \sin(2\cdot[\varphi - \gamma_2]).
	\label{eq:intensity}
\end{equation}
Here $I_0$ is a constant mean intensity onto which two modulations with the fundamental and double period are superimposed. The amplitudes and relative phases of two oscillations are denoted $I_{1,2}$ and $\gamma_{1,2}$ with indices $1$ and $2$ corresponding to one and two maxima per $2\pi$, respectively. Examples of the obtained best fits are plotted as solid lines in Fig. \ref{fig:results_3D} (b) reproducing the experimental data.
\begin{figure}[htb]
	\begin{center}
		\includegraphics[width=0.95\columnwidth]{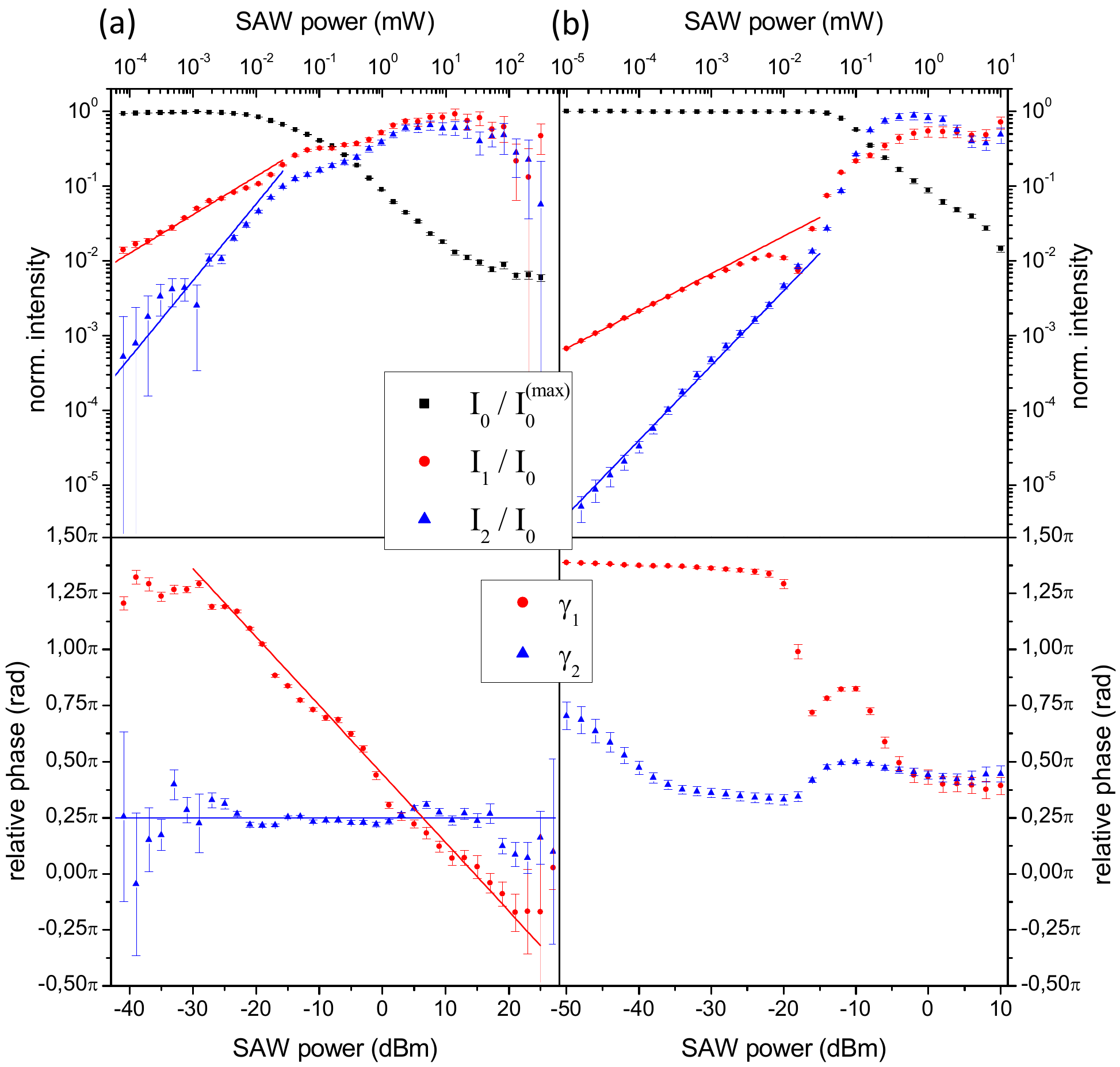}
		\caption{{\bf Figure 4} -- Fitted parameters of Eq. \eqref{eq:intensity} to the experimental data (a) and results of numerical simulations (b). The normalized amplitudes ($I_0$, $I_1$, $I_2$) and relative phases ($\gamma_1$, $\gamma_2$) are plotted as symbols in the upper and lower panels, respectively. The lines in the upper panels correspond to power laws with exponents 0.5 (red) and 1 (blue), the lines in the lower left panel are guides to the eye.}
		\label{fig:results_fit}
	\end{center}
\end{figure}
From the fits of Eq. \eqref{eq:intensity} to the experimental data we extracted the amplitudes $I_{1,2}$ and the relative phases $\gamma_{1,2}$ which are summarized in Fig. \ref{fig:results_fit} (a). In the upper panel, the intensities $I_1$ (red symbols), $I_2$ (blue symbols) and $I_0$ (black symbols) are plotted on a logarithmic scale as a function of the applied RF power. Here, $I_1$ and $I_2$ are normalized to $I_0$ at the respective RF power and $I_0$ to its maximum total intensity $I^{max}_0$. The constant offset $I_0$ shows the characteristic continuous quenching behavior as $P_{RF}$ is increased. At low RF powers, the amplitudes of the two modulations $I_1$ and $I_2$ show dissimilar variations as $P_{RF}$ is tuned. While $I_1$ follows a square root dependence $\propto\sqrt{P_{RF}}$, $I_2$ increases linearly as indicated by the two lines of slope 0.5 (red) and 1 (blue). The electric field $F$ induced by the SAW is proportional to its amplitude, scaling $A_{SAW}\propto\sqrt{P_{RF}}$. Thus, $I_1$ and $I_2$ increase linearly and quadratically with $F$. Since the drift velocity of $e$'s and $h$'s is $v_{e/h}=\mu_{e/h}\cdot F$, the linear and quadratic dependencies directly reflect the fact that one carrier species ($e$'s) or both carrier species ($e$'s \emph{and} $h$'s) are drifting in the SAW generated electric field. At high SAW power levels, both processes contribute equally and the two modulation amplitudes $I_1$ and $I_2$ become comparable to the mean signal level $I_0$. The analysis of the relative phases $\gamma_1$ and $\gamma_2$ as a function of $P_{RF}$ plotted in the lower panel of Fig. \ref{fig:results_fit} (a) shows dissimilar behavior for these two parameters. $\gamma_2$ is constant over the entire range of RF powers since this modulation marks the absolute phases of the stable points for $e$'s and $h$'s. In contrast, $\gamma_1$ exhibits a characteristic shift by $\sim\pi$ in the transition region between the mobility limited regime at small $P_{RF}<-25$\,dBm and the field-dominated regime $P_{RF}>10$\,dBm. These experimentally observed behaviors of the amplitudes $I_{0,1,2}$ and the relative phases $\gamma_{1,2}$ are well reproduced by our numerical simulations by using the same fitting procedure which was applied to the experimental data. The results of these fits are plotted in Fig. \ref{fig:results_fit} (b). Clearly, the calculated intensities (upper panel) the relative phases (lower panel) reproduce well the experimental data. The characteristic power laws for $I_1$ and $I_2$, constant value of $\gamma_2$ and the $\pi$ phase shift of $\gamma_1$ are clearly resolved in our simulations. 

In summary, we performed a combined experimental and theoretical study on the dynamic acousto-electric modulation of emission of a QW by a SAW. We investigated the underlying exciton dissociation and charge conveyance as a function of the applied acoustic power and phase of the SAW. We observe a characteristic transition of the SAW modulation from one emission maximum per SAW cycle at low acoustic powers to two maxima per SAW cycle at high power levels. At low acoustic powers transport of $h$'s is suppressed due to their low mobility compared to that of $e$'s, while at high acoustic powers, field-driven dissociation of $e$-$h$-pairs and charge conveyance of both $e$'s \emph{and} $h$'s occurs. From the measured modulation we are able to calibrate the absolute local phase of the SAW which is a crucial input parameter for the modeling of SAW-induced carrier injection into quantum dot nanostructures \cite{Voelk:2011,Voelk:2010}.\\

This work was supported via the Cluster of Excellence {\it Nanosystems Initiative Munich} (NIM) and the DFG via SFB 631 (TP B5) and the Emmy Noether Program (KR3790/2-1).

\end{document}